\documentclass[aps,prd,nofootinbib,floatfix,superscriptaddress,secnumarabic,preprintnumbers,longbibliography]{revtex4-2}
\usepackage{amsmath}
\usepackage{amsfonts}
\usepackage{graphicx}
\usepackage{subcaption}
\usepackage{xcolor}
\usepackage{commath}
\usepackage{hyperref}
\usepackage{cancel}
\usepackage{comment}
\usepackage[normalem]{ulem}

\newcommand{\MSbar}{{\overline{\rm MS}}}
\newcommand{\pa}{\partial}

\newcommand{\la}{\lambda}

\newcommand{\be}{\begin{equation}}
\newcommand{\ee}{\end{equation}}
\newcommand{\bea}{\begin{eqnarray}}
\newcommand{\eea}{\end{eqnarray}}

\begin{document}

\title{${\cal N}{=}1$ Supersymmetric QCD on the lattice using overlap fermions}

\author{M.~Costa}
\email[]{marios.costa@cut.ac.cy}
\affiliation{Present address: Department of Mechanical Engineering and Material Science and Engineering, Cyprus University of Technology, Limassol, CY-3036, Cyprus}
\affiliation{Department of Physics, University of Cyprus, Nicosia, CY-1678, Cyprus}
\affiliation{Rinnoco Ltd, Limassol, CY-3047, Cyprus}

\author{E.~Ioannou}
\email[]{ioannou.c.eleni@ucy.ac.cy}
\affiliation{Department of Physics, University of Cyprus, Nicosia, CY-1678, Cyprus}

\author{H.~Panagopoulos}
\email[]{panagopoulos.haris@ucy.ac.cy}
\affiliation{Department of Physics, University of Cyprus, Nicosia, CY-1678, Cyprus}

\begin{abstract}
Using ${\cal N}{=}1$ Supersymmetric QCD (SQCD) as a prototype model, this work presents a formulation of overlap quarks and gluinos on the lattice, with particular emphasis on the construction of chirally symmetric Yukawa terms. By incorporating the Ginsparg-Wilson relation, chiral transformations, and the Majorana condition for gluinos, we construct a consistent framework that preserves a lattice-modified chiral symmetry and reduces the number of required counterterms compared to Wilson-type discretizations. The formulation introduces auxiliary fermionic fields to realize exact chiral symmetry in Yukawa interactions and enables a detailed analysis of the resulting matrix structures. Upon functionally integrating out the auxiliary fields, ultralocal interaction terms emerge as new contributions to the lattice action. This approach provides a robust foundation for nonperturbative lattice studies of supersymmetric gauge theories. Future work will focus on computing all perturbative fine-tunings required in this formulation to enable continuum matching and numerical simulations of SQCD.
\end{abstract}

\maketitle

\section{Introduction}

Lattice formulations of Quantum Field Theories (QFTs) have been instrumental in the nonperturbative study of strong interactions, providing a well-defined framework~\cite{Wilson:1974sk} for investigating gauge theories beyond perturbation theory. In particular, overlap fermions implement an exact lattice realization of chiral symmetry via the Ginsparg-Wilson relation~\cite{Ginsparg:1981bj}, making them especially valuable for studying theories with massless fermions. This exact chiral symmetry on the lattice prevents additive mass renormalization for fermions and simplifies the renormalization of conserved currents, thereby reducing the number of fine-tuned parameters required in numerical simulations of QFTs. Although the overlap fermion action is not ultra-local, it avoids the fermion doubling problem while preserving the locality of the theory, in accordance with the constraints of the Nielsen–Ninomiya theorem~\cite{Nielsen:1981hk}.

Formulating a supersymmetric QFT on the lattice presents several challenges~\cite{Feo:2002yi, Giedt:2006pd, Catterall:2009it, Bergner:2009vg}, particularly in preserving supersymmetry (SUSY) while maintaining gauge invariance. A particular challenge, related to the inclusion of gluinos, squarks and quarks, regards the explicit breaking of chiral symmetry in Wilson-type fermion formulations. In this work, we examine the formulation of the ${\cal N}{=}1$ Supersymmetric QCD (SQCD) action, with $N_f$ quark flavors and color group $SU(N_c)$, using the overlap Dirac operator in the adjoint representation for the gluino fields and in the fundamental representation for the quark fields. We also incorporate quark mass terms in order to investigate their impact on the restoration of supersymmetry in lattice formulations of SQCD. Our analysis focuses on the extent to which the overlap formulation, which satisfies the Ginsparg-Wilson relation and suppresses chiral symmetry breaking, can improve the realization of supersymmetry on the lattice.

The Wilson formulation of SQCD on the lattice introduces several undesirable features that complicate the restoration of supersymmetry in the continuum limit. The Wilson term explicitly breaks chiral symmetry, leading to mixing between the squark components $A_+$ and $A_-$, defined in Section~\ref{cont}, which must be disentangled through finite renormalization. In addition, both quark and squark fields receive additive mass renormalizations which are power-divergent, requiring a problematic fine-tuning of their respective critical masses. A critical mass term for the gluino also emerges. Moreover, a second Yukawa term appears at the quantum level, that is not chirally invariant, and ten independent quartic scalar couplings are generated, some of which lack a chirally invariant structure. Given the rather generic nature of these complications, SQCD constitutes a highly valuable prototype for lattice studies of supersymmetric theories. It captures the essential structural elements and fine-tuning requirements present in more general non-chiral SUSY models, including gauge, matter, and scalar interactions. As such, it provides a practical and informative testbed for developing and validating lattice formulations of supersymmetry.

One key advantage of the overlap formulation is the suppression of some explicit SUSY-breaking terms induced by the Wilson term. Unlike Wilson fermions, where the massless limit requires intricate fine-tuning, overlap fermions naturally retain chiral symmetry at finite lattice spacing, reducing the number of required counterterms~\cite{Giedt:2009yd}. This property makes them particularly well-suited for simulations of SQCD with minimal explicit SUSY violations. Furthermore, overlap fermions naturally accommodate topological effects, which may play a crucial role in exploring vacuum structures in supersymmetric gauge theories.

This paper is organized as follows: In Section~\ref{cont} we present the continuum action for SQCD, while in Section~\ref{latt} we implement the overlap action in its lattice formulation. Subsection~\ref{SECyukawa} presents the construction of a chirally invariant Yukawa term on the lattice, respecting the representation structure of the color group $SU(N_c)$, and preserving the lattice-modified chiral symmetry. Subsections~\ref{SECNf1} and \ref{SECperturbative} study in more detail the one-flavor case and the perturbative setup, respectively. Lastly, in Section~\ref{summary} we summarize our findings and discuss future investigations of this formulation of the lattice action.

\section{Supersymmetric QCD in the continuum}
\label{cont}

In the Wess-Zumino (WZ) gauge~\cite{Wess:1974tw, deWit:1975veh}, the SQCD action contains the following fields: the gluon together with the gluino; and for each quark flavor, a Dirac fermion (quark) and two squarks. In the following we briefly define our notation. Although the action of SQCD used in this calculation can be found in literature, e.g. in Refs.~\cite{Wess:1992cp, Martin:1997ns, Costa:2017rht}, we present it here for completeness' sake; in the continuum and in Minkowski space, the action of SQCD is:
\bea
{\cal S}_{\rm SQCD}  =  \int d^4x \Big[ &-&\frac{1}{4}u_{\mu \nu}^{\alpha} {u^{\mu \nu}}^{\alpha} + \frac{i}{2} \bar \lambda^{\alpha}  \gamma^\mu {\cal{D}}_\mu\lambda^{\alpha}  \nonumber \\
&-& {\cal{D}}_\mu A_+^{\dagger}\,{\cal{D}}^\mu A_+ - {\cal{D}}_\mu A_- \,{\cal{D}}^\mu A_-^{\dagger}+ i \bar \psi  \gamma^\mu {\cal{D}}_\mu \psi  \nonumber \\
&-&i \sqrt2 g \big( A^{\dagger}_+ \bar{\lambda}^{\alpha}  T^{\alpha} P_+ \psi   -  \bar{\psi}  P_- \lambda^{\alpha}   T^{\alpha} A_+ +  A_- \bar{\lambda}^{\alpha}  T^{\alpha} P_- \psi   -  \bar{\psi}  P_+ \lambda^{\alpha}   T^{\alpha} A_-^{\dagger}\big)\nonumber\\  
&-& \frac{1}{2} g^2 (A^{\dagger}_+ T^{\alpha} A_+ -  A_- T^{\alpha} A^{\dagger}_-)^2 + m ( \bar \psi  \psi  - m A^{\dagger}_+ A_+  - m A_- A^{\dagger}_-)\Big] \,,
\label{susylagr}
\eea
where $\psi$ ($u_\mu$) is the quark (gluon) field, $u_{\mu \nu}$ is the gluon field tensor: $u_{\mu \nu} = \pa_{\mu}u_{\nu} - \pa_{\nu}u_{\mu} + i g\, [u_{\mu},u_{\nu}]$, $\lambda$ is the gluino field and $A_\pm$ are the squark complex scalar fields\footnote{Gauge symmetry requires the two squark fields $(A_+, A_-^\dagger)$ to lie in the fundamental representation of the color group $SU(N_c)$; their conjugates $(A_+^\dagger, A_-)$ are in the anti-fundamental representation.}; $T^{\alpha}$ are the generators of the $SU(N_c)$ gauge group in the fundamental representation, and satisfy the commutation relation:
\be
[T^\alpha, T^\beta] = i f^{\alpha\beta\gamma} T^\gamma,
\ee
where $f^{\alpha\beta\gamma}$ are the structure constants of the Lie algebra. The normalization is chosen such that:
\be
\text{Tr}(T^\alpha T^\beta) = \frac{1}{2} \delta^{\alpha\beta},
\ee
$P_\pm$ are projectors: $P_\pm= (1 \pm \,\gamma_5)/2$. Besides an implicit color index, quark and squark fields (as well as their masses, $m$) carry also an implicit flavor index; a summation over repeated indices is intended\footnote{Note that the first parenthesis in the last line of Eq.~(\ref{susylagr}) has an implicit double summation over independent flavor indices.}. The definitions of the covariant derivatives ${\cal{D}}_\mu$ are:
\bea
{\cal{D}}_\mu A_+ &=&  \pa_{\mu} A_+ + i g\,u_{\mu}^{\alpha}\,T^{\alpha}\,A_+, \nonumber \\
{\cal{D}}_\mu A_-^{\dagger} &=&  \pa_{\mu} A_-^{\dagger} + i g\,u_{\mu}^{\alpha}\,T^{\alpha}\,A_-^{\dagger}, \nonumber \\
{\cal{D}}_\mu A_- &=&  \pa_{\mu} A_- - i g\,A_-\,T^{\alpha}\,u_{\mu}^{\alpha}, \nonumber \\
{\cal{D}}_\mu A_+^{\dagger} &=&  \pa_{\mu} A_+^{\dagger} - i g\,A_+^{\dagger}T^{\alpha}\,u_{\mu}^{\alpha}, \nonumber \\
{\cal{D}}_\mu \psi &=&  \pa_{\mu} \psi+ i g\,u_{\mu}^{\alpha} \,T^{\alpha}\,\psi, \nonumber\\
{\cal{D}}_\mu \lambda &=&  \pa_{\mu} \lambda + i g \,[u_{\mu},\lambda]. 
\eea
The above action is invariant under the supersymmetric transformations ($\xi$ is a Grassmann Majorana spinor):
\bea
\delta_\xi A_+ & = & - \sqrt2  \bar\xi  P_+\psi  \, , \nonumber \\
\delta_\xi A_- & = & - \sqrt2  \bar\psi  P_+ \xi   \, , \nonumber \\
\delta_\xi (P_+ \psi) & = & i \sqrt2 ({\cal{D}}_\mu A_+) P_+ \gamma^\mu \xi   - \sqrt2 m P_+ \xi  A_-^{\dagger}\, , \nonumber \\
\delta_\xi (P_- \psi ) & = &  i \sqrt2 ({\cal{D}}_\mu A_-)^{\dagger} P_- \gamma^\mu \xi   - \sqrt2 m  A_+ P_- \xi \, ,\nonumber \\
\delta_\xi u_\mu^{\alpha} & = & -i \bar \xi  \gamma^\mu \lambda^{\alpha} , \nonumber \\
\delta_\xi \lambda^{\alpha}  & = & \frac{1}{4} u_{\mu \nu}^{\alpha} [\gamma^{\mu},\gamma^{\nu}] \xi  - i g \gamma^5 \xi  (A^{\dagger}_+ T^{\alpha} A_+ -  A_- T^{\alpha} A^{\dagger}_-)\,.
\label{susytransfDirac}
\eea

As in the case with the quantization of ordinary gauge theories, additional infinities will appear upon functionally integrating over gauge orbits. The standard remedy is to introduce a gauge-fixing term in the Lagrangian, along with a compensating Faddeev-Popov ghost term. The resulting Lagrangian, though no longer manifestly gauge invariant, is still invariant under Becchi-Rouet-Stora-Tyutin (BRST) transformations. This procedure of gauge fixing guarantees that Green's functions of gauge invariant objects will be gauge independent to all orders in perturbation theory. We use the ordinary gauge fixing term and ghost contribution arising from the Faddeev-Popov gauge fixing procedure\footnote{In what follows, the letter ``$\alpha$'', appearing as a superscript, stands for a color index in the adjoint representation, not to be confused with the gauge fixing parameter ``$\alpha$''.}:
\begin{equation}
{\cal S}^E_{GF}= \frac{1}{\alpha}\int d^4x {\rm{Tr}} \left( \partial_\mu u_\mu\right)^2,
\label{sgf}
\end{equation}
where $\alpha$ is the gauge parameter [\,$\alpha=1(0)$ corresponds to Feynman (Landau) gauge\,], and 
\begin{equation}
{\cal S}^E_{Ghost}= - 2 \int d^4x {\rm{Tr}} \left( \bar{c}\, \partial_{\mu}D_\mu  c\right), 
\label{sghost}
\end{equation}
where the ghost field $c$ is a Grassmann scalar which transforms in the adjoint representation of the gauge group, and: ${\cal{D}}_\mu c =  \pa_{\mu} c + i g \,[u_\mu,c]$. This gauge fixing term breaks supersymmetry. However, given that the renormalized theory does not depend on the choice of a gauge fixing term, and given that all known regularizations, in particular the lattice regularization, violate supersymmetry at intermediate steps, one may choose this standard covariant gauge fixing term instead of a supersymmetric one~\cite{Miller:1983pg}.

In Refs.~\cite{Costa:2017rht} and~\cite{Costa:2018mvb}, first lattice perturbative computations in the context of SQCD were presented; there, apart from the Yukawa and quartic couplings, we extracted the renormalization of all parameters and fields appearing in Eq.~(\ref{susylagr}) using Wilson gluons and fermions. In addition, we explored the mixing of some composite operators under renormalization. The results in these references~\cite{Costa:2017rht,Costa:2018mvb} will find further use in the present work. Furthermore, in the Wilson formulation we have calculated the fine-tunings of the Yukawa and quartic couplings (Refs.~\cite{Costa:2023cqv} and \cite{Costa:2024tyz}, respectively).

\section{Supersymmetric QCD on the lattice: Overlap Formulation}
\label{latt}

In our previous lattice calculation~\cite{Costa:2017rht, Costa:2018mvb, Costa:2023cqv, Costa:2024tyz}, we extended Wilson's formulation of the QCD action, to encompass SUSY partner fields as well. In this standard discretization quarks, squarks and gluinos are defined on the lattice sites, while gluons are defined on the links of the lattice: $U_\mu (x) = \exp\bigl(i g a T^{\alpha} u_\mu^\alpha (x+a\hat{\mu}/2)\bigr)$; $a$ is the lattice spacing. This formulation leaves no SUSY generators intact, and it also breaks chiral symmetry; thus, the need for fine-tuning will arise in numerical simulations of SQCD.  

To restore chiral symmetry at finite lattice spacing and reduce the extent of necessary fine-tuning, we now employ the overlap lattice formulation, which was introduced by Neuberger \cite{Neuberger:1997fp, Neuberger:1998wv} and preserves an exact lattice chiral symmetry through the Ginsparg-Wilson relation \cite{Ginsparg:1981bj}, eliminating additive mass renormalization for fermion fields and simplifying the renormalization of axial and supercurrent operators. Additionally, overlap fermions improve the continuum extrapolation by significantly reducing discretization artifacts that are present in Wilson-type formulations.

The Ginsparg–Wilson relation, which preserves a modified form of chiral symmetry on the lattice~\cite{Ginsparg:1981bj,Luscher:1998pqa} is:
\begin{equation} 
\gamma_5 D + D \gamma_5 = a D \gamma_5 D.
\label{GWrelation}
\end{equation}
This identity ensures that a lattice-modified chiral symmetry is maintained, which plays a crucial role in formulations involving Majorana fermions and supersymmetric theories.

Following the formulation in Ref.~\cite{Luscher:1998du}, the chiral transformations for quark fields are defined as:
\begin{equation} 
\delta \psi = \gamma_5 \left(1 - \frac{1}{2} a D \right) \psi, \quad 
\delta \bar{\psi} = \bar{\psi} \left(1 - \frac{1}{2} a D \right) \gamma_5,
\label{transformations} 
\end{equation}
where $D$ is the overlap Dirac operator. This transformation, elaborated further in~\cite{Clancy:2023ino}, ensures a well-defined and chirally symmetric lattice formulation of the theory even in the presence of Majorana fermions (e.g, gluino fields). Majorana fermions satisfy the property:
\begin{equation}
(\bar{\lambda}^\alpha)^T = \mathcal{C}\, \lambda^\alpha,
\end{equation}
where $\mathcal{C}$ is the charge conjugation matrix, which fulfills the relations: 
\be
(\gamma^\mu)^T \mathcal{C}^T = \mathcal{C} \gamma^\mu, \quad \mathcal{C}^{\dagger}\mathcal{C}=1, \quad \mathcal{C}^T = -\mathcal{C} \quad \Rightarrow \quad \mathcal{C} \gamma^5 = (\gamma^5)^T \mathcal{C}.
\label{Cproperties}\ee

In the case of quark fields, the inclusion of mass terms in the overlap formulation requires careful treatment, as discussed in Refs.~\cite{Niedermayer:1998bi,Alexandrou:1999wr, Alexandrou:2000kj}. In particular, by using the Ginsparg--Wilson relation in Eq.~(\ref{GWrelation}), one can derive the following identities involving the overlap Dirac operator $D$ and the chiral projectors:

\begin{equation}
D \, \frac{1 \pm \gamma_5 (1 - a D)}{2} = P_\mp D, \qquad
D \gamma_5 \left(1 - \frac{a}{2} D \right) + \left(1 - \frac{a}{2} D \right) \gamma_5 D = 0.
\label{otheridentity}
\end{equation}

These identities suggest a natural definition of the mass term in the lattice action. Instead of the naive term $m_0 \, \bar{\psi} \psi$, one considers the lattice-modified version:
\begin{equation}
m_0 \, \bar{\psi} \psi \quad \longrightarrow \quad m_0\, \bar{\psi} \left(1 - \frac{a}{2} D\right) \psi,
\end{equation}
such that the chiral transformation in Eq.~(\ref{transformations}) reproduces the correct continuum limit. The same structure as in the continuum theory confirms the consistency of the modified mass term with lattice chiral symmetry. Therefore, the mass term is introduced in the action as shown below, where the most common discretization of the Ginsparg--Wilson operator is given by the overlap formulation, $D = \mathcal{D}_{\text{ov}}$ (see Eq.~\ref{quarkoverlap}). It is important to emphasize the difference between chirality in the continuum and on the lattice. In the continuum, chirality is a local concept, independent of the gauge field. Both chiral rotations and projections involve only the spinor field at a single point $x$. On the lattice, however, the chiral rotation and projection operators involve the Ginsparg–Wilson operator $D$, which connects different  lattice sites. Consequently, lattice chirality depends on both the gauge field and information from neighboring points, making it inherently non-ultralocal. The quark lattice action takes the following form:
 
\begin{equation}
a^4 \sum_{x\,y} \bar{\psi}(x) \left( (1 - \frac{1}{2} a m_0) \mathcal{D}_{\text{ov}}(x,y)+ m_0 \delta_{x,y} \right) \psi(y).
\label{quark}
\end{equation}
By rescaling the fields, we define the bare mass:
\begin{equation}
m = \frac{m_0}{1 - \frac{1}{2} a m_0}.
\end{equation}

We observe that this expression can be rewritten as:
\begin{equation}
\bar{\psi}(1 - \tfrac{1}{2} a m_0) \left( \mathcal{D}_{\text{ov}} + \frac{m_0}{1 - \tfrac{1}{2} a m_0} \right) \psi,
\end{equation}
which has the same structure as a canonical massive Dirac action:
\begin{equation}
\bar{\psi}' (\mathcal{D}_{\text{ov}} - m) \psi',
\end{equation}
provided we make the identification:
\begin{equation}
-m = \frac{m_0}{1 - \tfrac{1}{2} a m_0}, \qquad \psi' = (1 - \tfrac{1}{2} a m_0)^{1/2} \, \psi.
\end{equation}

The overlap operator in the fundamental representation is defined as:
\begin{equation}
\mathcal{D}_{\text{ov}}(x,y) = \frac{1}{a} \left( \delta_{x,y} + \gamma_5 \frac{H(x,y)}{\sqrt{H(x,y)^\dagger H(x,y)}} \right) \,, \quad
H = \gamma_5 D_W \,,
\label{quarkoverlap}
\end{equation}
where $D_W$ is the Wilson--Dirac operator in the fundamental representation:
\begin{equation}
D_W = \frac{1}{2} \sum_\mu \left[ \gamma_\mu \left( \nabla_\mu + \nabla^*_\mu \right) - a\, \nabla^*_\mu \nabla_\mu \right] \,.
\end{equation}

The forward and backward covariant derivatives are given by: 
\begin{align}
\nabla_\mu(x,y) =  \frac{1}{a} \left( U_\mu(x) \, \delta_{x+a\hat\mu,y} - \openone \delta_{x,y} \right)\Rightarrow (\nabla_\mu\psi)(x) \equiv \sum_y \nabla_\mu(x,y)\psi(y) &= \frac{1}{a} \left[ U_\mu(x)\, \psi(x + a \hat{\mu}) - \psi(x) \right] \,, \\
\nabla^*_\mu(x,y) =  \frac{1}{a} \left(\openone \delta_{x,y} -  U^\dagger_\mu(x - a \hat{\mu}) \, \delta_{x- a \hat\mu,y} \right)\Rightarrow (\nabla^*_\mu\psi)(x) \equiv \sum_y \nabla^*_\mu(x,y)\psi(y) &= \frac{1}{a} \left[ \psi(x) - U^\dagger_\mu(x - a \hat{\mu})\, \psi(x - a \hat{\mu}) \right] \,.
\end{align}

For gluinos, given their Majorana nature, the corresponding action is given by:
\begin{equation} \label{Gluino_action}
- \ a^4 \sum_{x\,y} {\rm{Tr}}\bigl(\la^T(x)\, \mathcal{C} \,\mathcal{D}_{\text{ov}}^{\text{adj}}(x,y) \la(y)\bigr),
\end{equation}
which aligns with the small-momentum behavior of the continuum theory~\cite{Clancy:2023ino}. The definition of the overlap operator in the adjoint reprensentation $\mathcal{D}^{\text{adj}}_{\text{ov}}$ takes the form~\cite{Clancy:2023ino}:
\begin{equation} \label{gluino_overlap_operator}
    \mathcal{D}^{\text{adj}}_{\text{ov}}=\frac{\mu}{2}(1+V_{\text{maj}}),
\end{equation}
where $\mu$ is a constant with dimensions of mass that behaves as $\sim1/a$, and $V_{\text{maj}}$ is defined as:
\begin{equation}
    V_{\text{maj}}=D^{\text{adj}}_W(D_W^{\text{adj}\dagger} D_W^{\text{adj}})^{-1/2}, 
\end{equation}
where \( D_W^{\text{adj}} \) is the Wilson--Dirac operator in the adjoint representation:
\begin{equation}
   D_W^{\text{adj}}=-\mu+\frac{1}{2}\sum_{\mu}\left[\gamma_{\mu}(\nabla_{\mu}^{\text{adj}}+\nabla_{\mu}^{*\text{adj}})-\mu a^2\nabla_{\mu}^{*\text{adj}}\nabla_{\mu}^{\text{adj}}\right].
\end{equation}

One can easily show that the operator $V_{\text{maj}}$ satisfies the relations: 
\begin{equation}
    V_{\text{maj}}^{\dagger}V_{\text{maj}}=1, \qquad
    \mathcal{C}V_{\text{maj}}\mathcal{C}^{-1}=V_{\text{maj}}^T,\qquad
    \gamma_5V_{\text{maj}}\gamma_5=V_{\text{maj}}^\dagger.
\end{equation}

The kinetic term for the gluino in the action (Eq.~\ref{Gluino_action}) is invariant under the following infinitesimal chiral transformation:
\begin{equation}
    \delta\lambda=\gamma_5\left(1-\frac{1}{2}(1+V_{\text{maj}})\right)\lambda, \quad  \quad \delta\bar\lambda=\bar\lambda\left(1-\frac{1}{2}(1+V_{\text{maj}})\right)\gamma_5
\end{equation}
The covariant derivatives in the adjoint representation are:
\begin{align}
\nabla^{\text{adj}}_\mu(x,y) =  \frac{1}{a} \left( U^{\text{adj}}_\mu(x)\, \delta_{x+a\hat\mu,y} - \openone \, \delta_{x,y} \right) \Rightarrow \nabla^{\text{adj}}_\mu \la^\alpha(x) &= \frac{1}{a} \left[ \bigl(U^{\text{adj}}_\mu(x)\bigr)^{\alpha \beta} \la^\beta(x + a \hat{\mu}) - \la^\alpha(x) \right] \,, \\
\nabla^{*\,\text{adj}}_\mu(x,y) =  \frac{1}{a} \left(\openone \,\delta_{x,y} - U^{\text{adj}\,\dagger}_\mu(x - a \hat{\mu})\, \delta_{x-a \hat\mu,y}  \right) \Rightarrow \nabla^{*\,\text{adj}}_\mu \la^\alpha(x) &= \frac{1}{a} \left[ \la^\alpha(x) - \bigl(U^{\text{adj}\,\dagger}_\mu(x - a\hat{\mu})\bigr)^{\alpha \beta} \la^\beta(x - a\hat{\mu}) \right] \,,
\end{align}
where the adjoint link variables are constructed from the fundamental ones as:
\begin{equation}
\left[ U_\mu^{\text{adj}}(x) \right]^{\alpha \beta} = 2\, \mathrm{Tr} \left( T^\alpha\, U_\mu(x)\, T^\beta\, U^\dagger_\mu(x) \right) \,.
\end{equation}
From the above, it follows that:
\begin{equation}
\nabla^{\text{adj}}_\mu \la(x) = \frac{1}{a} \left[ U_\mu(x) \la(x + a \hat{\mu})U_\mu^\dagger(x) - \la(x) \right]\,.
\end{equation}
Henceforth, the arguments $(x,y)$ will be omitted whenever this does not lead to ambiguity.

For the overlap discretization, the Euclidean lattice action ${\cal S}^{L}_{\rm SQCD}$ takes the form shown below. We note that, in its current form, the Yukawa interaction term (third line in the following expression) is not chirally invariant; this issue will be addressed in the following subsection. 
\begin{align}
{\cal S}^{L}_{\rm SQCD} =\; & a^4 \sum_x \Bigg\{
\frac{N_c}{g^2} \sum_{\mu,\nu} \left( 1 - \frac{1}{N_c}\, \mathrm{Tr} \, U_{\mu\nu}(x) \right) - \sum_y \mathrm{Tr} \left( \lambda^T(x)\, \mathcal{C}\, \mathcal{D}_{\text{ov}}^{\text{adj}}(x,y)\, \lambda(y) \right) \nonumber \\
&+ \mathcal{D}_\mu A^\dagger_+(x)  \mathcal{D}_\mu A_+(x)
+ \mathcal{D}_\mu A_-(x)  \mathcal{D}_\mu A^\dagger_-(x) + \sum_y \bar{\psi}(x) \left[ \left(1 - \frac{a m_0}{2} \right) \mathcal{D}_{\text{ov}}(x,y) + m_0 \delta_{x,y} \right] \psi(y) \nonumber \\
&+ i \sqrt{2} g \Big(
A_+^\dagger(x)\, \bar{\lambda}^\alpha(x)\, T^\alpha P_+ \psi(x)
- \bar{\psi}(x)\, P_- \lambda^\alpha(x)\, T^\alpha A_+(x)  + A_-(x)\, \bar{\lambda}^\alpha(x)\, T^\alpha P_- \psi(x)
- \bar{\psi}(x)\, P_+ \lambda^\alpha(x)\, T^\alpha A_-^\dagger(x)
\Big) \nonumber \\
&+ \frac{1}{2} g^2 \left( A_+^\dagger(x)\, T^\alpha A_+(x) 
- A_-(x)\, T^\alpha A_-^\dagger(x) \right)^2 + m^2 \left( A_+^\dagger(x)\, A_+(x) + A_-(x)\, A_-^\dagger(x) \right)
\Bigg\} \,,
\label{susylagrLattice}
\end{align}
where: $U_{\mu \nu}(x) =U_\mu(x)U_\nu(x+a\hat\mu)U^\dagger_\mu(x+a\hat\nu)U_\nu^\dagger(x)$, and summations over flavors are implicit in the last three lines of Eq.~(\ref{susylagrLattice}).  The definitions of the lattice covariant derivatives on squark fields are as follows:
\bea
\label{DAplus}
{\cal{D}}_\mu A_+(x) &\equiv& \frac{1}{a} \Big[  U_\mu (x) A_+(x + a \hat{\mu}) - A_+(x)  \Big],\\
{\cal{D}}_\mu A_+^{\dagger}(x) &\equiv& \frac{1}{a} \Big[A_+^{\dagger}(x + a \hat{\mu}) U_\mu^{\dagger}(x)  -  A_+^\dagger(x)\Big],\\
{\cal{D}}_\mu A_-(x) &\equiv& \frac{1}{a} \Big[A_-(x + a \hat{\mu}) U_\mu^{\dagger}(x)  -  A_-(x)\Big],\\
\label{DAminusdagger}
{\cal{D}}_\mu A_-^{\dagger}(x) &\equiv& \frac{1}{a} \Big[U_\mu (x) A_-^{\dagger}(x + a \hat{\mu})   -  A_-^{\dagger}(x) \Big].
\eea
In Eqs.~(\ref{DAplus})–(\ref{DAminusdagger}), we avoid using the symmetric derivative—not only to limit the interactions to at most two lattice points, but more importantly, to avoid ``scalar doubling", which might otherwise necessitate the introduction of a Wilson term. We note, however, that the symmetries of the action remain the same regardless of the choice between both types of derivatives. 

For a perturbative treatment, a gauge-fixing term $\mathcal{S}^E_{\text{GF}}$, together with the compensating ghost field term $\mathcal{S}^E_{\text{Ghost}}$, must also be added to the action, just as in the continuum, in order to avoid divergences from the integration over gauge orbits; these terms are the same as in the non-supersymmetric case. Furthermore, a standard ``measure'' term $\mathcal{S}^E_{\text{Meas.}}$ must be added to the action, in order to account for the Jacobian in the change of integration variables: $U_\mu \to u_\mu$\,. All the details and definitions of $\mathcal{S}^E_{\text{GF}}$, $\mathcal{S}^E_{\text{Ghost}}$, $\mathcal{S}^E_{\text{Meas.}}$ can be found in Ref.~\cite{rothe2012} (see also~\cite{Costa:2017rht}).

\subsection{Chirally invariant Yukawa term}
\label{SECyukawa}

To maintain exact lattice chiral symmetry and properly formulate supersymmetric Yukawa interactions using overlap fermions, we adapt Lüscher’s prescription~\cite{Luscher:1998pqa} and introduce auxiliary fermion fields:
\begin{itemize}
    \item $\chi_\lambda^\alpha(x)$: auxiliary Majorana fermion in the adjoint representation of $SU(N_c)$, \\ \phantom{$\chi_\lambda^\alpha(x)$:\,} corresponding to the gluino field,
    \item $\chi_\psi(x)$: auxiliary Dirac fermion (with implicit flavor index) in the fundamental representation of $SU(N_c)$, \\ \phantom{$\chi_\lambda^\alpha(x)$:\,} corresponding to the quark field.
\end{itemize}
These fields are non-dynamical and vanish in the continuum limit, but are essential for writing lattice chirally invariant Yukawa terms.

\medskip

Before addressing the Yukawa term, the auxiliary field contribution to the Euclidean lattice action is:
\begin{equation}
\mathcal{S}_{\text{aux}} = a^4 \sum_x \left( 
- \mu \, \bar{\chi}_\lambda^\alpha(x)\, \chi_\lambda^\alpha(x)
- \frac{2}{a} \, \bar{\chi}_\psi(x)\, \chi_\psi(x) 
\right),
\label{auxiliary}
\end{equation}
where $\mu$ denotes the same parameter introduced in Eq.~(\ref{gluino_overlap_operator}), and summation over color and flavor indices is implied. A modified chiral transformation is introduced, in which the auxiliary fields transform under lattice chiral symmetry in such a way that their sum with the corresponding quark and gluino fields satisfy the standard continuum chiral transformation. Explicitly, for the quark field $\psi$, we define the infinitesimal modified chiral transformations~\cite{Luscher:1998pqa}:
\bea
\label{modified_transformations1}
\delta \psi = \gamma_5 \left(1 - \frac{1}{2} a D \right) \psi + \gamma_5 \chi_\psi, &\quad& 
\delta  \bar{\psi} = \bar{\psi} \left(1 - \frac{1}{2} a D \right) \gamma_5 + \bar\chi_{\psi} \gamma_5,
 \\
\delta \chi_\psi = \gamma_5 \frac{1}{2} a D \psi, &\quad& 
\delta  \bar\chi_{\psi} = \bar{\psi} \frac{1}{2} a D \gamma_5,
\label{modified_transformations2} 
\eea
so that the combinations $(\psi + \chi_\psi)$ and $(\bar \psi + \bar \chi_\psi)$ transform as:
\begin{equation}
\delta (\psi + \chi_\psi) = \gamma_5 (\psi + \chi_\psi), \qquad
\delta (\bar \psi + \bar \chi_\psi) = (\bar \psi + \bar \chi_\psi)\, \gamma_5,
\end{equation}
Similarly, for the gluino field $\lambda$~\cite{Clancy:2023ino}:
\bea
\label{modified_transformations3}
\delta\lambda=\gamma_5\left(1-\frac{1}{2}(1+V_{\text{maj}})\right)\lambda + \gamma_5\chi_\lambda,  &\quad&  
\delta\bar\lambda=\bar\lambda\left(1-\frac{1}{2}(1+V_{\text{maj}})\right)\gamma_5+\bar\chi_\lambda\gamma_5,
\\
\delta\chi_\lambda=\gamma_5\frac{1}{2}(1+V_{\text{maj}})\lambda,  &\quad&
\delta\bar\chi_\lambda=\bar\lambda\frac{1}{2}(1+V_{\text{maj}})\gamma_5,
\label{modified_transformations4} 
\eea
and the combinations ($\lambda + \chi_\lambda$) and ($\bar\lambda + \bar \chi_\lambda$) transform as:
\begin{equation}
    \delta(\lambda+\chi_\lambda)=\gamma_5(\lambda+\chi_\lambda), \qquad
     \delta(\bar\lambda+\bar\chi_\lambda)=(\bar\lambda+\bar\chi_\lambda)\gamma_5.
\end{equation}
The modified chiral transformations in Eqs.~(\ref{modified_transformations1}),~(\ref{modified_transformations2}),~(\ref{modified_transformations3}) and (\ref{modified_transformations4}) not only ensure that the combinations $(\psi + \chi_\psi)$ and $(\lambda + \chi_\lambda)$ transform in the standard continuum chiral manner, but also leave invariant the modified Yukawa interaction in Eq.~(\ref{auxiliary-Yukawa}). Further, the sum of the second term in $\mathcal{S}_{\text{aux}}$ and the kinetic term for quarks in Eq.~(\ref{quark}) with $m_0 \to 0$ is also invariant. The same statement holds for the first term in $\mathcal{S}_{\text{aux}}$ and the kinetic term for gluinos in Eq.~(\ref{Gluino_action}). 

\medskip

The Yukawa part of the action (third line of Eq.~\ref{susylagrLattice}) now takes the chirally invariant form:
\begin{align}
\mathcal{S}_{\text{Y}} =\;  a^4 \sum_x \, i &\sqrt{2} g \Big[ 
A_+^\dagger(x) \left( \bar{\lambda}^\alpha(x) + \bar{\chi}_\lambda^\alpha(x) \right) T^\alpha P_+ \left( \psi(x) + \chi_\psi(x) \right)- \left( \bar{\psi}(x) + \bar{\chi}_\psi(x) \right) P_- \left( \lambda^\alpha(x) + \chi_\lambda^\alpha(x) \right) T^\alpha A_+(x) \nonumber \\
& \quad + A_-(x) \left( \bar{\lambda}^\alpha(x) + \bar{\chi}_\lambda^\alpha(x) \right) T^\alpha P_- \left( \psi(x) + \chi_\psi(x) \right) - \left( \bar{\psi}(x) + \bar{\chi}_\psi(x) \right) P_+ \left( \lambda^\alpha(x) + \chi_\lambda^\alpha(x) \right) T^\alpha A_-^\dagger(x)
\Big].
\label{auxiliary-Yukawa}
\end{align}
Since the auxiliary fields $\chi_\psi$ and $\chi_\lambda^\alpha$ are introduced in order to restore exact lattice chiral symmetry via a modified chiral transformation, and they appear both quadratically and linearly in the lattice action, they can be functionally integrated out exactly (quantum treatment). This procedure yields auxiliary-field-independent contributions in the form of determinants (or Pfaffians in the case of Majorana fields), along with exponential factors, which will be explicitly described below. The resulting effective action involves only the physical fermionic degrees of freedom, $\psi$ and $\lambda^\alpha$, as well as the squark fields, $A_+$ and $A_-$. The total lattice action remains invariant under a modified chiral symmetry and is consistent with the Ginsparg--Wilson relation. This formulation ensures the correct continuum limit while preserving exact chiral symmetry at finite lattice spacing.

\medskip
Overall, the auxiliary fields appear in  $\mathcal{S}_{\text{Y}}$ and $\mathcal{S}_{\text{aux}}$\,. A classical treatment of these fields would amount to inserting in the action the expressions for $\chi_\psi$ and $\chi_\lambda$ which result from solving their classical equations of motion:
\bea
- \frac{2}{a} \chi_\psi 
&=& i \sqrt{2} g T^\alpha (P_- A_+ + P_+ A_-^\dagger) (\la^\alpha + \chi_\lambda^\alpha), \label{EOM1}\\
\frac{2}{a} \bar{\chi}_\psi^T 
&=&  i \sqrt{2} g (T^\alpha)^T (A_+^\dagger P_+ + A_- P_-) \, \mathcal{C}\,  (\la^\alpha + \chi_\la^\alpha), \label{EOM2}
\eea

\vspace{-0.65cm}

\begin{flalign}
\text{and:}\qquad
2 \mu \, \mathcal{C}\,\chi^\alpha_\lambda 
= i \sqrt{2} g \left[ \left( A_+^\dagger \, \mathcal{C}\, T^\alpha P_+ + A_- \, \mathcal{C}\,  T^\alpha P_- \right) (\psi+ \chi_\psi) 
- \left( A_+ (T^\alpha)^T P_- + A_-^\dagger (T^\alpha)^T P_+ \right) (\bar{\psi}^T + \bar{\chi}_\psi^T) \right]. &&
\label{EOM3}
\end{flalign}
Eqs. (\ref{EOM1}, \ref{EOM2}, \ref{EOM3}) are a coupled set of linear algebraic equations, consistent with the Majorana condition.

\medskip
For a quantum treatment of the auxiliary fields $\chi_\psi$ and $\chi_\lambda^\alpha$\,, we must functionally integrate over both. Given that the first field is a Dirac spinor, while the second is Majorana, it is more convenient to carry out the two integrations in sequence, starting with $\chi_\psi$\,.

The part of the action containing $\chi_\psi$ is (see Eqs.~(\ref{auxiliary}, \ref{auxiliary-Yukawa}); we omit the common argument $x$ in all fields):
\begin{align}
\mathcal{S}_{\chi_\psi} = a^4 \sum_x \Bigl( 
- \frac{2}{a} \, \bar{\chi}_\psi\, \chi_\psi
+ i \sqrt{2}  & g \Big[ 
A_+^\dagger \bigl( \bar{\lambda}^\alpha + \bar{\chi}_\lambda^\alpha \bigr) T^\alpha P_+ \chi_\psi -   \bar{\chi}_\psi P_- \bigl( \lambda^\alpha + \chi_\lambda^\alpha \bigr) T^\alpha A_+ \nonumber \\
& + A_- \bigl( \bar{\lambda}^\alpha + \bar{\chi}_\lambda^\alpha \bigr) T^\alpha P_- \chi_\psi  - \bar{\chi}_\psi P_+ \bigl( \lambda^\alpha + \chi_\lambda^\alpha \bigr) T^\alpha A_-^\dagger
\Big]
\Bigr),
\label{S-chi-psi}
\end{align}

Employing the standard Gaussian functional integration over fermions:
\be
\int \mathcal{D}\bar{X} \, \mathcal{D}X \; \exp\left[ -\int d^4x \left( \bar{X} \mathcal{M} X + \bar{X} \Psi + \bar{\Psi} X \right) \right]
= \det(\mathcal{M}) \; \exp\left( \int d^4x\,  \bar{\Psi} \, \mathcal{M}^{-1} \Psi \right),\label{Integral_X}
\ee
in its lattice discretized form, we obtain:
\be
\int \! \mathcal{D}\bar{\chi}_\psi  \mathcal{D}\chi_\psi  \exp( - \mathcal{S}_{\chi_\psi})
= \bigl(\frac{2}{a}\bigr)^{4N_c N_f V}  \exp\Bigl( -a^4 \sum_x a g^2 \bigl( \bar{\lambda}^\alpha + \bar{\chi}_\lambda^\alpha \bigr)\bigl[(A_+^\dagger T^\alpha T^\beta A_-^\dagger)P_+ + (A_- T^\alpha T^\beta A_+)P_- \bigr] \bigl( \lambda^\beta + \chi_\lambda^\beta \bigr) \Bigr),\label{Integral-chi-psi}
\ee
where $V$ is the number of lattice sites; the constant prefactor multiplying the exponential can be dropped, while the exponent provides an effective contribution to the total SQCD action.

To integrate now over the auxiliary field $\chi_\lambda^\alpha$, let us first single out the parts of the action containing this field (including contributions from Eq.~(\ref{Integral-chi-psi}))\,:
\begin{align}
\mathcal{S}_{\chi_\lambda} = a^4 \sum_x \Bigl( 
&- \mu \, \bar{\chi}_\lambda^\alpha\, \chi_\lambda^\alpha
+ i \sqrt{2} g \Big[ 
A_+^\dagger \bar{\chi}_\lambda^\alpha T^\alpha P_+ \psi 
- \bar{\psi} P_- \chi_\lambda^\alpha T^\alpha A_+ 
+ A_- \bar{\chi}_\lambda^\alpha T^\alpha P_- \psi  
- \bar{\psi} P_+ \chi_\lambda^\alpha  T^\alpha A_-^\dagger
\Big]
\nonumber\\
& + \frac{a\,g^2}{2}\, \big[
\bar{\chi}_\lambda^\alpha Q^{\alpha\beta}\chi_\lambda^\beta 
+ \bar{\chi}_\lambda^\alpha Q^{\alpha\beta}\lambda^\beta 
+ \bar{\lambda}^\alpha Q^{\alpha\beta}\chi_\lambda^\beta 
\big] \Bigr),
\end{align}
\begin{flalign}
\text{where:} &\quad Q^{\alpha\beta}\equiv 
(A_+^\dagger \{T^\alpha, T^\beta\} A_-^\dagger) P_+ 
+ (A_- \{T^\alpha, T^\beta\} A_+) P_- \, . &
\end{flalign}
\begin{align}
\Rightarrow \mathcal{S}_{\chi_\lambda} &= a^4 \sum_x \Bigl(
\bar{\chi}_\lambda^\alpha 
\bigl[-\mu\, \delta^{\alpha\beta}+ \frac{a\,g^2}{2} Q^{\alpha\beta} \bigr] 
\chi_\lambda^\beta 
\nonumber\\
& \hspace{1.5cm} 
+ \bar{\chi}_\lambda^\alpha \bigl[i \sqrt{2} g (A_+^\dagger  P_+ + A_- P_-) T^\alpha \psi 
+ \frac{a\, g^2}{2} Q^{\alpha\beta} \lambda^\beta \bigr]
\nonumber \\
& \hspace{2.0cm} 
- \bigl[ i \sqrt{2} g \bar{\psi} T^\beta (A_+ P_- + A_-^\dagger P_+) 
- \frac{a\, g^2}{2} \bar{\lambda}^\alpha Q^{\alpha\beta} \bigr]
\chi_\lambda^\beta
\Bigr) 
\nonumber\\
&\equiv a^4 \sum_x \Bigl(
\bar{\chi}_\lambda^\alpha \,\Xi^{\alpha\beta} \chi_\lambda^\beta 
+ \bar{\chi}_\lambda^\alpha\, \Psi^\alpha 
+ \bar{\Psi}^\alpha\, \chi_\lambda^\alpha 
\Bigr) ,
\label{S-chi-lambda}
\end{align}
where only the anticommutator $\{T^\alpha,T^\beta\}$ contributes to $Q^{\alpha\beta}$, due to the Majorana nature of $(\lambda^\alpha + \chi_\lambda^\alpha)$. 

[Note: For definiteness, we will use $\gamma_5^T = \gamma_5$\,, as is the case in all commonly used bases of $\gamma$ matrices. There exist other bases, of course, which are compatible with Eq.~(\ref{Cproperties}) and in which $\gamma_5$ is antisymmetric, such as: $\gamma_0 = \openone \otimes \sigma_1 ,\ \gamma_i = -\sigma_i \otimes \sigma_3 ,\ \gamma_5 = \openone \otimes \sigma_2,\ \mathcal{C} = \gamma_2$\,. By Eq.~(\ref{Cproperties}), a symmetric $\gamma_5$ commutes with $\mathcal{C}$.]

The functional integral for a generic Majorana fermion $\lambda$ and a generic antisymmetric matric $\mathcal{M}$ reads:
\begin{equation}
\int \mathcal{D}\lambda  \; \exp\left[ -\int d^4x \Bigl( \lambda^T \,\mathcal{M} \,\lambda + \bar{\lambda} \eta + \bar{\eta} \lambda \Bigr) \right]
= \text{\text{Pf}}(2\mathcal{M}) \; \exp\left( -\frac{1}{4}\int d^4x\,  (\bar{\eta}-\eta^T \mathcal{C}) \, \mathcal{M}^{-1} (\bar{\eta}^T +\mathcal{C} \eta)\right),\label{Integral_Majorana}
\end{equation}
where $\text{\text{Pf}}(2\mathcal{M})$ stands for the Pfaffian of $2\mathcal{M}$. Integration over $\chi_\lambda$ thus leads to (see Eq.~(\ref{S-chi-lambda}) for the definitions of $\Xi$, $\Psi$ and $\bar{\Psi}$):
\begin{align}
\int \mathcal{D}\chi_\lambda  \; \exp( -\mathcal{S}_{\chi_\lambda})
&= \text{\text{Pf}}(2\,\mathcal{C}^T\, \Xi) \; \exp\left( -\frac{1}{4}\, a^4 \sum_x  \Bigl(\bar{\Psi}^\alpha-(\Psi^T)^\alpha \mathcal{C}\Bigr) \, \bigl((\mathcal{C}^T \,\Xi)^{-1}\bigr)^{\alpha\beta} \Bigl((\bar{\Psi}^T)^\beta +\mathcal{C} \Psi^\beta\Bigr)\right) \nonumber\\
&\equiv \text{Pf}(2\,\mathcal{C}^T\, \Xi) \; \exp\left( - \mathcal{S}_{\Psi\Psi}\right).\label{Integral-chi-lambda}
\end{align}

 An important characteristic of the  quadratic part in $\mathcal{S}_{\chi_\lambda}$ is that it is purely local, i.e. the matrix $\Xi$ is strictly diagonal in coordinate space. A consequence of this is that the inverse matrix appearing in the Gaussian integration over auxiliary fields is a tensor sum of the inverses of $\Xi$ at each space-time point separately. Similarly, the Pfaffian stemming from the integration is actually a product of Pfaffians at each space-time point. Thus, in all cases, integration over the auxiliary fields will only amount to introduction of the additional {\it ultralocal} contribution, $\mathcal{S}_{\Psi\Psi}$, in the action, as well as the Pfaffian which can also be expressed as an ultralocal contribution (see Eq.~(\ref{SPf})). A salient feature of $\mathcal{S}_{\Psi\Psi}$ is the appearance of ``Majorana-type" lattice artifacts, of the form $\psi\psi$ and $\bar\psi\bar\psi$, for the quark fields.

 We can simplify $\text{Pf}(2\,\mathcal{C}^T\, \Xi)$ by evaluating it in a basis\footnote{While the Pfaffian does not share all the basis-invariant properties of a determinant, it is certainly invariant under all similarity transformations which are symmetries of Eq.~(\ref{Integral_Majorana}); these include orthogonal transformations and, in particular, a basis change from Dirac to Weyl fermions.} where:
 \be
 \gamma_5 = \openone \otimes \sigma_3, \qquad \mathcal{C} = -i \sigma_2 \otimes \openone \,.
 \ee
 Writing:
 \be
 \Xi \equiv \Xi_+ \, P_+ + \Xi_- \, P_- \qquad \Bigl[\, (\Xi_+)^{\alpha\beta} = -\mu \, \delta^{\alpha\beta} + \frac{a\,g^2}{2} \, A_+^\dagger \{T^\alpha , T^\beta\} A_-^\dagger \,,\quad
 (\Xi_-)^{\alpha\beta} = -\mu \, \delta^{\alpha\beta} + \frac{a\,g^2}{2} \, A_- \{T^\alpha , T^\beta\} A_+ \, \Bigr]
 \label{Xi}
 \ee
we obtain:
\be
2\,\mathcal{C}^T\, \Xi = 2
\begin{pmatrix}
i\, \sigma_2 \, \Xi_+ & 0 \\ 0 & i\, \sigma_2 \, \Xi_-
\end{pmatrix}\quad
\Rightarrow \quad \text{Pf}(2\,\mathcal{C}^T\, \Xi) = 2^{2(N_c^2-1)} \cdot \det(\Xi_+) \cdot \det(\Xi_-)
\label{Pfaffian}\ee
Similarly, the inverse of $\Xi$, appearing in Eq.~(\ref{Integral-chi-lambda}), simplifies to:
\be
\Xi^{-1} = (\Xi_+)^{-1} \, P_+ + (\Xi_-)^{-1} \, P_-
\ee
Being ultralocal, $\Xi_+$ and $\Xi_-$ are $(N_c^2-1)\times (N_c^2-1)$ matrices whose inverse and determinant can be easily evaluated numerically. Their perturbative value will be examined in Subsection~\ref{SECperturbative}.

\medskip
In conclusion, after integration over the auxiliary fields, the action of SQCD acquires three additional contributions: 
\begin{itemize}
    \item The first contribution, stemming from Eq.~(\ref{Integral-chi-psi}), is:
\be 
\mathcal{S}_{\lambda\lambda} = a^4 \sum_x\frac{a \, g^2}{2} \, \bar{\lambda}^\alpha \bigl[(A_+^\dagger \{T^\alpha, T^\beta\} A_-^\dagger)P_+ + (A_- \{T^\alpha, T^\beta\} A_+)P_- \bigr] \lambda^\beta 
\label{Sll}\ee  

    \item The Pfaffian in Eq.~(\ref{Integral-chi-lambda}) (see also Eqs.~(\ref{Xi}, \ref{Pfaffian})) can be cast as a second contribution to the action, by taking its logarithm:
\be
\text{Pf}(2\,\mathcal{C}^T\, \Xi) \equiv \exp(-\mathcal{S}_{\text{Pf}}), \qquad \mathcal{S}_{\text{Pf}} =-\text{tr}\log(\Xi_+) - \text{tr}\log(\Xi_-) + \text{constant} \label{SPf}
\ee

    \item The third contribution is $\mathcal{S}_{\Psi\Psi}$, shown in Eq.~(\ref{Integral-chi-lambda}); to make it more explicit:

\begin{align}
&\mathcal{S}_{\Psi\Psi} = \frac{1}{4}\, a^4 \sum_x  \Bigl(\bar{\Psi}^\alpha-(\Psi^T)^\alpha \mathcal{C}\Bigr) \, \bigl((\mathcal{C}^T \,\Xi)^{-1}\bigr)^{\alpha\beta} \Bigl((\bar{\Psi}^T)^\beta +\mathcal{C} \Psi^\beta\Bigr),
\label{SPsiPsi} \\
\text{where: }\quad & \Psi^\alpha = \phantom{-}i \sqrt{2} g (A_+^\dagger  P_+ + A_- P_-) T^\alpha \psi + \frac{a\, g^2}{2} Q^{\alpha\beta} \lambda^\beta,  \quad \bar\Psi^\alpha = -i \sqrt{2} g \,\bar{\psi} T^\alpha (A_+ P_- + A_-^\dagger P_+) + \frac{a\, g^2}{2} \bar{\lambda}^\beta Q^{\beta\alpha} \nonumber\\
& Q^{\alpha\beta}= (A_+^\dagger \{T^\alpha, T^\beta\} A_-^\dagger) P_+ + (A_- \{T^\alpha, T^\beta\} A_+) P_- , \quad \Xi^{\alpha\beta} = -\mu\, \delta^{\alpha\beta}+ \frac{a\,g^2}{2} \, Q^{\alpha\beta} .\nonumber
\end{align}
\end{itemize}
Thus, the action of SQCD in the overlap lattice formulation reads:
\be
{\cal S}^{\rm Total}_{\rm SQCD} = {\cal S}^{L}_{\rm SQCD} + \mathcal{S}_{\lambda\lambda} + \mathcal{S}_{\text{Pf}} \,+ \mathcal{S}_{\Psi\Psi} \,.
\ee

For a perturbative treatment, one also needs to add: $\mathcal{S}^E_{\text{GF}} + \mathcal{S}^E_{\text{Ghost}} +\mathcal{S}^E_{\text{Meas.}}$, as previously mentioned.

\subsection{The case $N_f=1$}
\label{SECNf1}
In order to shed some light on the characteristics of the matrices $\Xi_\pm$\,, let us focus on the one-flavor case, $N_f=1$. While in non-supersymmetric QCD this case is of limited phenomenological interest, in SQCD it elucidates the structure of poles arising from the elimination of the auxiliary fields.

What are the eigenvalues and eigenvectors of $\Xi_-$ (Eq.~(\ref{Xi}))? [The treatment of $\Xi_+$ will be identical, merely substituting $A_\pm$ by $A_\mp^\dagger$\,.]
One eigenvector, denoted $v_0$\,, is:
\begin{flalign}
\hspace{0.5cm} (v_0)^\beta = (A_-\,T^\beta A_+), &&
\text{with  eigenvalue: } e_0 = -\mu + \frac{a\,g^2}{2} \Bigl(1-\frac{1}{N_c}\Bigr) (A_- A_+).\hspace{2cm}
\label{v0}\end{flalign}
A set of degenerate eigenvectors, $v_-^\alpha$, are:
\begin{flalign}
\hspace{0.5cm}(v_-^\alpha)^\beta = (A_- [T^\alpha,T^\beta] A_+),&&
\text{with  eigenvalue: } e_- = -\mu + \frac{a\,g^2}{4} \, (A_- A_+).\hspace{3.4cm}
\label{v-}\end{flalign}
Finally, there is also a set of degenerate eigenvectors, $v_+^\alpha$\,:
\begin{flalign}
\hspace{0.5cm}(v_+^\alpha)^\beta = -\frac{1}{2} \Bigl(1-\frac{1}{N_c}\Bigr) (A_-A_+)^2 \delta^{\alpha\beta} + \Bigl(1-\frac{1}{N_c}\Bigr) (A_- A_+) (A_- \{T^\alpha,T^\beta\} A_+) -\Bigl(1-\frac{2}{N_c}\Bigr) (A_- T^\alpha A_+)(A_- T^\beta A_+) \,,&&
\nonumber\end{flalign}

\vspace{-0.8cm}

\begin{flalign}
&&\text{with  eigenvalue: } e_+ = -\mu \,.\hspace{5.65cm}
\label{v+}\end{flalign}

In order to find the multiplicities of the above eigenvectors, and to check that they are complete, we form the projectors $\Pi_0$, $\Pi_-$, $\Pi_+$ to the subspaces spanned by $v_0$, $v_-^\alpha$ and $v_+^\alpha$, respectively:
\be
\Pi_0^{\ \beta\gamma} \equiv \mathcal{N}_0 \,(v_0)^\beta (v_0)^\gamma ,\qquad
\Pi_-^{\ \beta\gamma} \equiv \mathcal{N}_- \,(v_-^\alpha)^\beta (v_-^\alpha)^\gamma ,\qquad
\Pi_+^{\ \beta\gamma} \equiv \mathcal{N}_+ \,(v_+^\alpha)^\beta (v_+^\alpha)^\gamma ,
\ee
where the values of the normalization coefficients $\mathcal{N}_0$, $\mathcal{N}_-$, $\mathcal{N}_+$,  are such that: $(\Pi_0)^2 = \Pi_0$, $(\Pi_-)^2 = \Pi_-$, $(\Pi_+)^2 = \Pi_+$, respectively. Using the expressions in Eqs.~(\ref{v0}), (\ref{v-}) and (\ref{v+}) for the eigenvectors, we find:
\bea
\Pi_0^{\ \beta\gamma} &=& 2 \Bigl(1-\frac{1}{N_c}\Bigr)^{-1} (A_- A_+)^{-2} (A_- T^\beta A_+)\,
(A_- T^\gamma A_+)\nonumber\\
\Pi_-^{\ \beta\gamma} &=& 2 (A_- A_+)^{-1} (A_- \{T^\beta,T^\gamma\} A_+) - 4 (A_- A_+)^{-2}\,(A_- T^\beta A_+)
(A_- T^\gamma A_+)\nonumber\\
\Pi_+^{\ \beta\gamma} &=& \delta^{\beta\gamma} - 2 (A_- A_+)^{-1} (A_- \{T^\beta,T^\gamma\} A_+)
+ 2 \Bigl(1-\frac{1}{N_c}\Bigr)^{-1} \Bigl(1 - \frac{2}{N_c}\Bigr) (A_- A_+)^{-2} (A_- T^\beta A_+)\,
(A_- T^\gamma A_+)
\eea

There follows immediately that the above eigenvectors are a complete and mutually orthogonal set, since:
\be
\Pi_0 + \Pi_- + \Pi_+ = \openone\,, 
\ee
\be
(v_0)^\gamma (v_-^\alpha)^\gamma = 0, \qquad
(v_0)^\gamma (v_+^\alpha)^\gamma = 0, \qquad
(v_-^\alpha)^\gamma (v_+^\beta)^\gamma = 0, \qquad \forall \alpha,\,\beta.
\ee

The multiplicity of the eigenvalues $e_0$\,, $e_-$, $e_+$ equals the rank of the corresponding projectors, respectively:
\be
\text{tr}(\Pi_0) = 1, \quad \text{tr}(\Pi_-) = 2 \, (N_c -1), \quad \text{tr}(\Pi_+) = N_c^2 - 2 N_c\,.
\ee
It is straightforward to verify that $\text{tr}(\Xi_-)$ equals the sum over all of its eigenvalues, multiplied by their respective multiplicities:
\be
e_0 \, \text{tr}(\Pi_0) + e_- \, \text{tr}(\Pi_-) + e_+ \, \text{tr}(\Pi_+) = -\mu \, (N_c^2 -1) + a\, g^2 \,\frac{N_c^2 -1}{2N_c} \,\bigl(A_- A_+\bigr) = \text{tr}(\Xi_-)
\ee

Thus, the determinant of $\Xi_-$ equals:
\be
\det (\Xi_-) = \Bigl(-\mu + \frac{a\,g^2}{2} \Bigl(1-\frac{1}{N_c}\Bigr) (A_- A_+)\Bigr) \Bigl(-\mu + \frac{a\,g^2}{4} \, (A_- A_+)\Bigr)^{2\,(N_c-1)} \bigl(-\mu)^{N_c^2 - 2 N_c}.
\label{detXi}\ee

The inverse of $\Xi_-$ is necessarily of the form:
\be
\bigl((\Xi_-)^{-1}\bigr)^{\beta\gamma} = c_1 \, \delta^{\beta\gamma} + c_2 \, (A_-\{T^\beta,T^\gamma\} A_+) + c_3 \, (A_- T^\beta A_+) (A_- T^\gamma A_+),
\ee
where the coefficients $c_i$ are given by:
\bea
c_1 &=& - \frac{1}{\mu} \nonumber \\
c_2 &=& \frac{a\,g^2}{2\mu} \, \Bigl(-\mu+ \frac{a\,g^2}{4} (A_- A_+)\Bigr)^{-1}\nonumber \\
c_3 &=& -\frac{a^2\,g^4}{4\mu}\, \Bigl(-\mu+ \frac{a\,g^2}{4} (A_- A_+)\Bigr)^{-1}\, \Bigl(-\mu+ \frac{a\,g^2}{2} (A_- A_+) \bigl(1-\frac{1}{N_c}\bigr)\Bigr)^{-1}\, \bigl(1-\frac{2}{N_c}\bigr)
\eea

We see that $(\Xi_-)^{-1}$ has poles, leading to potential divergences for large values of the squark fields $A_\pm$\,. In a perturbative treatment these poles bear no consequences, since $(\Xi_-)^{-1}$ can be expanded as a power series in $g^2$ with convergent coefficients. But even nonperturbatively, these poles are harmless, since they will be exactly compensated by corresponding zeros in the determinant of $\Xi_-$ (see Eq.~(\ref{detXi})); this is to be expected, since both features are consequences of a Grassmann integration (Eq.~(\ref{Integral-chi-lambda})) which is necessarily finite (see also Ref.~\cite{Luscher:1998pqa}).

\subsection{Perturbative setup}
\label{SECperturbative}
In this subsection we provide an expansion of the overlap action of SQCD in powers of the bare lattice coupling constant $g$, including terms up to order $g^5$. This expansion is sufficient for a perturbative study of SQCD up to two loops, in order to renormalize all elementary fields, the action parameters, and relevant composite operators.

A number of important perturbative checks can be performed on overlap SQCD, to verify that the overlap formulation evades some undesirable features of the more standard, chirality non-preserving formulations which employ Wilson-like gluinos and quarks. Such undesirable features are: Critical (power divergent) bare masses for quarks, squarks and gluinos; mixing of the $A_+$ and $A_-^\dagger$ squark fields; appearance of unwanted $\mathcal{O}(a^0)$ counterterms for both Yukawa and quartic couplings.

The perturbative treatment of overlap quarks in non-supersymmetric QCD, and of overlap gluinos in supersymmetric pure Yang-Mills theory, has been extensively addressed in the past (see, e.g., Refs.~\cite{Kikukawa:1998py,Constantinou:2007rm,Constantinou:2007gv}). Furthermore, the additional terms in the SQCD Lagrangian at the classical level (Eq.~(\ref{susylagrLattice})) coincide with the corresponding ones in the Wilson-fermion formulation of SQCD \cite{Costa:2017rht}. Thus, we only need to focus on the three new contributions to the Lagrangian, which arise as a result of the functional integration over the auxiliary fields $\chi^\alpha_\lambda,\ \chi_\psi$ and are shown in Eqs.~(\ref{Sll}, \ref{SPf}, \ref{SPsiPsi}). We recall that these auxiliary fields were introduced in order to preserve chiral invariance in the Yukawa part of the action, a novel issue which was not present either in QCD or in pure SUSY Yang-Mills.

\begin{itemize}
\item 
The first contribution, $\mathcal{S}_{\lambda\lambda}$\,, being a monomial in $g$, is already in a form suitable for perturbation theory.
\item
The second contribution, $\mathcal{S}_{\text{Pf}}$\,, expanded in powers of $g$ becomes: 
\bea
\mathcal{S}_{\text{Pf}} &=&  \frac{a\,g^2}{2\mu} \Bigl(\frac{N_c^2-1}{N_c}\Bigr) \Bigl((A_+^{\dagger f} A_-^{\dagger f})+(A_-^f A_+^f)\Bigr) \nonumber\\
&&+ \frac{a^2g^4}{16\mu^2} \Bigl[\Bigl((A_+^{\dagger f} A_-^{\dagger f})(A_+^{\dagger f'} A_-^{\dagger f'}) + (A_-^f A_+^f)(A_-^{f'} A_+^{f'})\Bigr)\Bigl(\frac{N_c^2+2}{N_c^2}\Bigr)\nonumber\\
&&\hspace{1.2cm} +\Bigl((A_+^{\dagger f} A_-^{\dagger f'})(A_+^{\dagger f'} A_-^{\dagger f}) + (A_-^f A_+^{f'})(A_-^{f'} A_+^f)\Bigr)\Bigl(\frac{N_c^2-4}{N_c}\Bigr)\Bigr] + \mathcal{O}(g^6).
\label{SPfperturbative}\eea
[Flavor indices are shown explicitly in Eq.~(\ref{SPfperturbative}), for the sake of clarity; summation over paired indices is implied.]
\item
The third contribution, $\mathcal{S}_{\Psi\Psi}$, is shown in Eq.~(\ref{SPsiPsi}). For its perturbative expression, modulo terms of $\mathcal{O}(g^6)$, it suffices to substitute $\Xi^{-1} = (\Xi_+)^{-1} \, P_+ + (\Xi_-)^{-1} \, P_-$ in $\mathcal{S}_{\Psi\Psi}$ by its expansion  to first subleading order:
\be
\bigl((\Xi_+)^{-1}\bigr)^{\alpha\beta} = -\frac{1}{\mu} \delta^{\alpha\beta} - \frac{a\,g^2}{2\mu^2} A_+^\dagger \{T^\alpha , T^\beta\} A_-^\dagger + \mathcal{O}(g^4)\,, \quad
\bigl((\Xi_-)^{-1}\bigr)^{\alpha\beta} = -\frac{1}{\mu} \delta^{\alpha\beta} - \frac{a\,g^2}{2\mu^2} A_- \{T^\alpha , T^\beta\} A_+ + \mathcal{O}(g^4)\,.
\ee
\end{itemize}

\begin{figure}[ht!]
    \centering
  \includegraphics[scale=0.55]{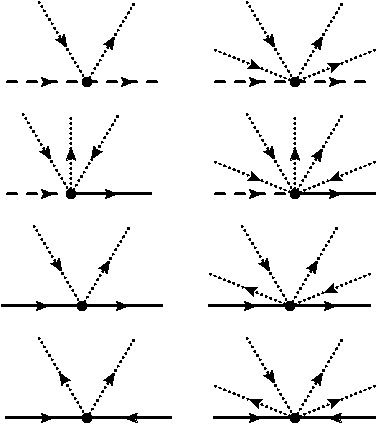}
    \caption{Feynman vertices derived from the additional contributions to the action, $\mathcal{S}_{\lambda\lambda}$, $\mathcal{S}_{\text{Pf}}$, $\mathcal{S}_{\Psi\Psi}$.
    Solid lines represent quark fields ($\psi$), dashed lines represent gluino fields ($\lambda$), and dotted lines denote squarks ($A_\pm$). 
    A squark line arrow entering (exiting) a vertex denotes an $A_+$ ($A_+^\dagger$) field; 
    the opposite is true for $A_-$ ($A_-^\dagger$) fields. 
    Vertices with reversed arrows are also present.}
    \label{fig:Leff_vertices}
\end{figure}

Inspection of the three contributions to the action (Eqs.~(\ref{Sll}, \ref{SPf}, \ref{SPsiPsi})) reveals a number of new interaction vertices which were not present in the lattice formulation with Wilson fermions. These vertices are higher order in $g$ and in $a$\,; they typically contain additional powers of squark-antisquark pairs, such as: $\bar\lambda{-}\lambda{-}A^\dagger{-}A$, $\bar\psi{-}\psi{-}A^\dagger{-}A$, $\bar\psi{-}A{-}\lambda{-}(A^\dagger A)^n$. In addition, there are Majorana-type quark couplings: $\psi{-}A^\dagger{-}\psi{-}A^\dagger$. Their representation as Feynman diagrams is shown in Fig.~\ref{fig:Leff_vertices}. 

\section{Summary -- Future plans}
\label{summary}

This work presents a consistent lattice formulation of ${\cal N}{=}1$ Supersymmetric QCD (SQCD) using overlap fermions, which preserve a modified form of chiral symmetry on the lattice through the Ginsparg–Wilson relation. In order to construct Yukawa terms that respect this symmetry, we have introduced appropriate auxiliary fields and we have obtained a chirally invariant lattice action that incorporates gluino, quark, and squark fields. Compatibly with the Nielsen-Ninomiya theorem, since the action is now local, but not ultra-local, this approach will not lead to any doublers, while leaving the theory (for vanishing mass) chirally invariant in a modified sense. Within this framework, we have developed an action suitable for both non-perturbative and perturbative investigations. This enables the computation of Green's functions, the renormalization of the theory, and the fine-tuning of bare parameters. This formulation naturally avoids additive mass renormalization and significantly reduces the number of supersymmetry-breaking counterterms required, in contrast to Wilson-type discretizations. To implement the integration over auxiliary fields, we perform a detailed matrix analysis for an arbitrary number of flavors, $N_f$, with explicit simplifications demonstrated in the single-flavor case $N_f=1$. Additionally, we carry out a perturbative expansion of the resulting expressions. 

A natural extension of this work involves the complete renormalization of the theory. Future efforts will focus on computing all necessary perturbative fine-tunings and counterterms on the lattice and in the $\MSbar$ scheme to enable precise matching with the continuum SQCD theory. We will retain the quark mass term  in this formulation throughout our calculations, as quark and squark masses are consistent with supersymmetry. The initial step is the renormalization of fields and masses, which will clarify whether the squark fields acquire critical masses, while quark and gluino fields remain protected from additive mass renormalization by the overlap formalism. This analysis will also determine whether the squark components mix among themselves and whether the second Yukawa term arises. These counterterms often present in Wilson-based formulations, are suppressed by chiral symmetry, indicating improved supersymmetry preservation. Furthermore, the renormalization of the four-squark couplings will reveal whether multiple independent quartic interactions emerge, as in the Wilson case. 

\begin{acknowledgments}
 The project is implemented under the programme of social cohesion``THALIA 2021-2027'' co-funded by the European Union through the Cyprus Research and Innovation Foundation (RIF). We would like to thank David Kaplan for very useful communication regarding his publication, Ref.~\cite{Clancy:2023ino}. 
\end{acknowledgments}

\end{document}